Impact of Declining Proposal Success Rates on Scientific Productivity

September 11, 2015

**AAAC Proposal Pressures Study Group**
Priscilla Cushman, Todd Hoeksema, Chryssa Kouveliotou, James Lowenthal, Bradley Peterson,
Keivan G. Stassun, Ted von Hippel

**Executive Summary:** Over the last decade proposal success rates in the fundamental sciences have dropped significantly. Astronomy and related fields funded by NASA and NSF are no exception. Data across agencies show that this is *not* principally the result of a decline in proposal merit (the proportion of proposals receiving high rankings is largely unchanged), nor of a shift in proposer demographics (seniority, gender, and institutional affiliation have all remained unchanged), nor of an increase (beyond inflation) in the average requested funding per proposal, nor of an increase in the number of proposals per investigator in any one year. Rather, the statistics are consistent with a scenario in which agency budgets for competed research are flat or decreasing in inflation-adjusted dollars, the overall population of investigators has grown, and a larger proportion of these investigators are resubmitting meritorious but unfunded proposals, likely in response to the decreased success rates. Recent research on the time cost of proposal writing versus that of producing publishable results show that a funding rate of ~6% represents the tipping point below which proposal writing prevents more papers than grants produce. This is close to the success rate experienced by new investigators against an overall average funding rate of 20%, due to rating bias against PIs without recent funding. At this 20% average selection threshold, the opportunity cost is still significant (2-3 papers per successful proposal) even for established researchers. Unfortunately, even an investigator submitting a proposal rated "very good" can expect, with three attempts, only a ~58% chance of funding. A 20% overall funding rate is thus unhealthy for the field, since it precludes stable, long-term support for students, postdocs, or researchers on soft money, and it preferentially discourages young researchers from remaining in the field. Yet, as demonstrated below, we are currently in exactly this situation. We conclude that an aspirational proposal success rate of 30-35% would still provide a healthily competitive environment for researchers, would more fully utilize the scientific capacity of the community's facilities and missions, and provide relief to the funding agencies who face the logistics of alarming volumes of proposals.

## 1. Introduction

Recent years have witnessed a significant decline in the fraction of successful proposals for basic research in science, engineering, and the health disciplines. In this study, which covers astronomy and astrophysics, the trend can be attributed to an increasing number of proposals submitted per funding opportunity, while the agency budgets are flat or declining in inflation-adjusted dollars. While this basic trend is incontrovertible, the fundamental cause(s) and, therefore, the appropriate response(s) have been less clear. The goal should be to sustain the U.S. astronomical research program at an acceptable and stable level, with an appropriately competitive funding environment, while maintaining the highest productivity and quality level commensurate with US leadership. Proposal writing should not prevent more science from being done than the sought-after funding would have enabled. To do this effectively, the astronomical community needs to determine the quantitative proposal success rate threshold that should be considered "too low", and then recommend and actively encourage policies that avoid dropping below that threshold.

This brief report from the AAAC Proposal Pressures Study Group presents the interim results of an ongoing analysis of funding agency statistics, together with findings from recent literature on the impact of proposal writing. Our purpose is to assess how declining success rates affect the health of research in the astronomical sciences. This is a direct response to the 2014 AAAC report[1] recommendation: "*The AAAC and the agencies should work together to clarify and quantify the questions related to individual*

---

[1] http://www.nsf.gov/mps/ast/aaac/reports/annual/aaac_2014_report.pdf

*investigator grants and mid-scale programs raised in this report. Other groups such as the American Astronomical Society and the National Research Council's Committee on Astronomy and Astrophysics should be involved as appropriate. The goal should be a clearer factual base on which to assess the health of the current individual investigator grants programs and make recommendations for future improvements."* The 2015 AAAC report[2] included preliminary findings from the newly formed study group, many of which are included here. In this short report, we also put the effect of falling success rates in context, by examining results from "A Survey Analysis of Grant Writing Costs and Benefits."[3]

## 2. Proposal success rates and demographic trends

The data clearly indicate that the number of proposals submitted to NASA and NSF for individual investigator and mid-scale grants in astronomy and astrophysics, planetary sciences, and heliophysics is increasing faster than the available funding, causing a corresponding drop in success rate. The data show that the PIs submitting these proposals have remained a stable demographic entity in terms of race, gender, number of years since PhD, type of institution, and number of proposals submitted per opportunity. However, as discussed below, the data do indicate that proposers are now more likely to resubmit their meritorious but unfunded proposals. The data therefore suggest that researchers consequently spend more time re-submitting their proposals, often to no avail. We consider the impact on scientific productivity in Section 3.

### 2.1 Funding in NSF and NASA

The NSF Division of Astronomical Sciences (NSF/AST) allocated $43.7M to its Astronomy and Astrophysics Grant (AAG) program in FY2014, about 17.5% of its total division budget of $239M. This fraction has held relatively stable over the last 10 years. Also stable is the relative breakdown of labor (50% of AAG grant allocations) vs. travel, overhead, and equipment. While budgets doubled during the 1990s, funds have just barely kept up with inflation on average over the last decade, and are expected to shrink over the next 5 years as NSF takes on operations costs for new facilities (e.g. DKIST+LSST) coming online. On the other hand, the number of AAG grant requests to NSF/AST has risen from 238 in 1990 to 732 in 2014, with most of the increase in the last decade. As a result, the grant success rate has fallen from 50% in 1990 to close to 15% currently. The data from the last decade are shown in Figure 1. Note that in the absence of any portfolio readjustment, and assuming current proposal submission trends and a budget that continues to be flat, the proposal success rate is projected to drop to ~10% in FY19. NSF/AST is presently moving forward rapidly to divest those optical and radio telescopes recommended by the Portfolio Review to prevent this scenario. If divestment proceeds as recommended by the Portfolio Review Committee, if the budget continues to be flat, and if modest trimming of proposal budgets is maintained at the levels practiced by NSF/AST in recent years, then success rates can be stabilized at no better than ~15%. Note that the NSF/AST funding rate was indeed a much healthier 30-35% in the early 2000's, such that investigators submitting strong (VG to E) proposals could expect a manageable risk of ~30% of no funding after three attempts as shown in section 2.

Another example within NSF is the Particle Astrophysics program in the Division of Physics, a much smaller program that studies astronomy and astrophysics with particles: cosmic rays, high-energy photons, neutrinos, and dark matter particles. Over the past decade, their budget has been a steady percentage of the Physics budget, around 7%. While the program is small compared to NSF/AST, a similar story emerges. From 2005 to 2014 the number of proposals more than doubled (from 30 to 70), whereas funding increased by about 34% in the same period. The average success rate from 2005 to 2007 was 45%, while from 2012 to 2014 the success rate had decreased to 39%.

Over the last decade, the budget for NASA Astrophysics Research Opportunities in Space and Earth Sciences programs (APRA, ADAP, ATP, XRP, WPS) has risen slowly from $71M to $80M, while the number of submitted proposals has increased much more rapidly, from around 500 in FY04 to around 800 in FY15. With some fluctuations, the number of selected proposals and the funding levels for selected proposals have remained approximately constant, so the success rate for proposals has decreased from ~30% to ~18% over the decade.[4] On a positive note, including the NASA Astrophysics Guest Observer programs increases the overall success rate to closer to 25%.

---

[2] http://www.nsf.gov/mps/ast/aaac/reports/annual/aaac_2015_report.pdf
[3] von Hippel and von Hippel (2015): http://journals.plos.org/plosone/article?id=10.1371/journal.pone.0118494

The solar and space physics community relies primarily on NASA Heliophysics (125 proposals per year) and the NSF Division of Atmospheric and Geospace Sciences (25 proposals per year in Solar-Terrestrial). Overall, inflation has steadily eroded the buying power of NASA Heliophysics research funding, which has remained essentially constant in real dollars since FY04. The success rate has decreased steadily over the last 8 years, dropping from ~35% to 17%.[5] The same story is generally true for NASA's Planetary Science Division. The total Planetary budget, adjusted for inflation and set to constant 2014 dollars, dropped from $1,731M (2004) to $1,380M (2015), with most of the drop occurring around FY13. Since 2004 the percent of Planetary proposals accepted dropped from above 30% to below 20%.

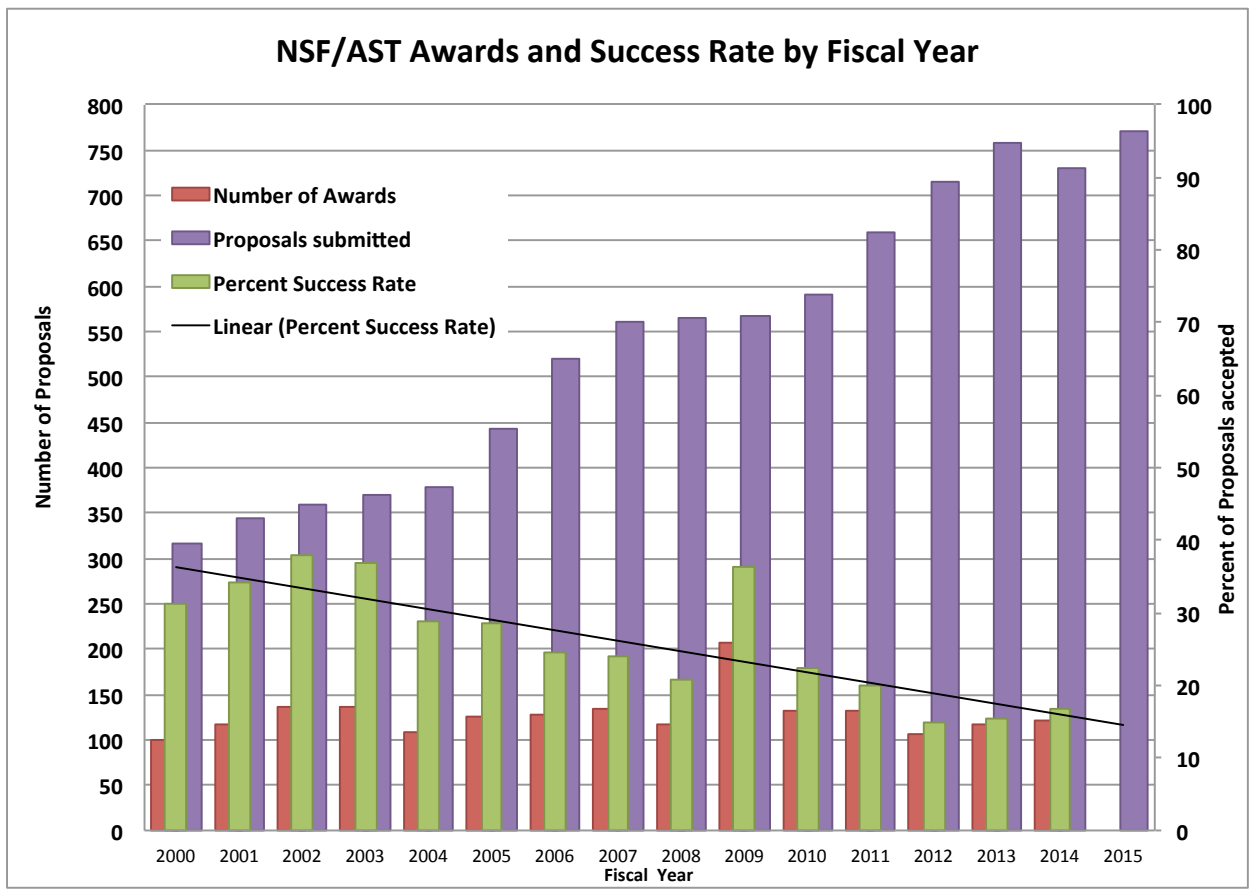

*Figure 1.* Historical NSF/AST (AAG) proposal success rate[6] through 2014. The anomalous spike in FY09 is due to the one-time stimulus provided by the American Recovery and Reinvestment Act.

---

[4] http://science.nasa.gov/media/medialibrary/2015/01/07/AAS_Townhall_Hertz_Jan2015_FINAL.pdf
[5] http://science.nasa.gov/researchers/sara/division-corner/heliophysics-division-corner/
[6] http://www.nsf.gov/attachments/131083/public/Dan-Evans_AST_Individual_Investigator_Programs-AAAC_Meeting.pdf

*2.2 Proposal quality*

NASA Astrophysics has tracked scores for many years and has reasonable confidence in the stability of the scale. Using NASA proposal selection data from all Science Mission Directorate ROSES programs[7] from 2007 to 2012, we see a pattern emerge. The proportion of all submitted proposals that are rated Very Good to Excellent (VG, VG/E, E) is roughly constant, with some evidence for at most a ~10% decrease. Also stable across all programs are the success rates for VG/E and E at >75% and >90%, respectively. However, the success rate of proposals in the VG category is rapidly falling from 45% in 2007-2008 to 25% in 2012. Thus, while the quality of submitted proposals remains high, a greatly increasing fraction of meritorious proposals can no longer be funded.

*2.3 Proposer demographics*

NSF data from the Astronomy and Astrophysics Grants (AAG) program show that the rise in proposals is driven largely by an increase in the number of investigators participating in a given year, each submitting on average one proposal. The proportion of individuals submitting two or more proposals has experienced only a modest rise, from 16% to 21%. Data from NASA do not have as straightforward a breakdown, but the same story emerges. For example, for NASA Astrophysics programs (ADAP, ATP, WPS and XRP) in 2014, there were 573 proposals from 476 unique PIs[8]. In 2008 and 2009, the number of proposals rose from 290 to 393. Thus we know that there were no more than 290 (393) unique PIs in 2008 (2009), compared to 476 in 2014. Note that none of these data can address whether PIs are submitting a larger number of similar proposals to multiple agencies; this information is not readily distilled.

The hypothesis that funding opportunities and fellowships targeting postdocs have created a proportionally larger population, just now moving into the ranks of PI, is also not borne out by the statistics. Indeed, the proportion of submitting PIs who are less than 15 years since PhD has actually declined somewhat in NSF AST, from ~50% in FY06 to ~45% in FY15. In NSF Particle Astrophysics, the fraction of younger PIs was very small[9] when the division program was created in 2000, so the uptick in younger researchers since then has simply brought it into the same balance observed by NSF/AST over the last decade.

There is also no significant change in the proportion of institution types submitting proposals, nor in the average salary-months requested by PIs, nor in the average requested inflation-adjusted budgets, nor in the relative success rates by gender and ethnicity. The data do not directly speak to possible changes in the proportion of "soft money" investigators driving up proposal pressure, however to be consistent with the above findings, such soft money investigators would have to comprise a mix of junior and senior investigators in such a way as to leave the overall demographics unchanged. It is thus clear that, while the number of proposers has been rising steadily, the demographic pool from which they are drawn has not significantly changed. The mix of junior and senior investigators is largely unchanged, and the number of investigators submitting multiple proposals has increased only modestly.

The number of unique proposers to NSF/AST AAG per 3-year funding cycle has grown from 1025 (2008-2010) to 1160 (2013-2015), representing an average growth of 2.2% per year when corrected for resubmission by unsuccessful PIs in subsequent years (detailed in section 2.4). This can be compared to the AAS full membership, which has grown from 3000 to 4500 over the years from 1990 to 2014, representing an average growth of 2.1% per year. Thus, it appears that growth in the number of investigators tracks the overall growth of researchers in the field.

---

[7] http://science.nasa.gov/researchers/sara/grant-stats/a-plot-of-grades-vs-who-gets-selected/
[8] http://science.nasa.gov/media/medialibrary/2014/04/09/2014.03.27_ApS_RA_final-2.pdf
[9] NSF/PHY Program in Particle Astrophysics data. Provided by J. Whitmore.

*2.4 Proposal resubmissions*

Figure 2 shows that the average number of unique proposers over a typical 3-year grant cycle (number in the right panel divided by 3) is less than the number proposing each year. These data (and anecdotal evidence) indicate that principal investigators are reapplying in subsequent years when their 3-year proposals are not funded. This has the effect of driving up the number of proposals each year, and driving down the success rates even further. While we do not have direct access to how many of the proposals are for new 3-year grants and how many are resubmissions following a failed proposal from the year before, we can create quantitative models under some general assumptions. For example, if we assume that the number of new proposals remains constant each year and that the falling success rates of Figure 1 are applied each year equally to the mix of new and repeat proposals, then the resubmission rate can be obtained. Under these conditions, a simple fit to the each-year and 3-year submission rate data in Figure 2 yields a 70% resubmission rate. To interpret this result, imagine that the number of new proposals and resubmitted proposals had been equal in 2008 (not unreasonable), then by 2014 the number of new proposals would be only 40% of all submitted proposals. Thus, resubmission in general has the secondary effect of driving up the number of proposals and driving down the success rates even further. Previously, a larger fraction of investigators could submit funding proposals roughly every 3 years, whereas now they increasingly (re)propose in consecutive years, for some (as yet unknown) number of years. We revisit the issue of repeat submissions and success rates below.

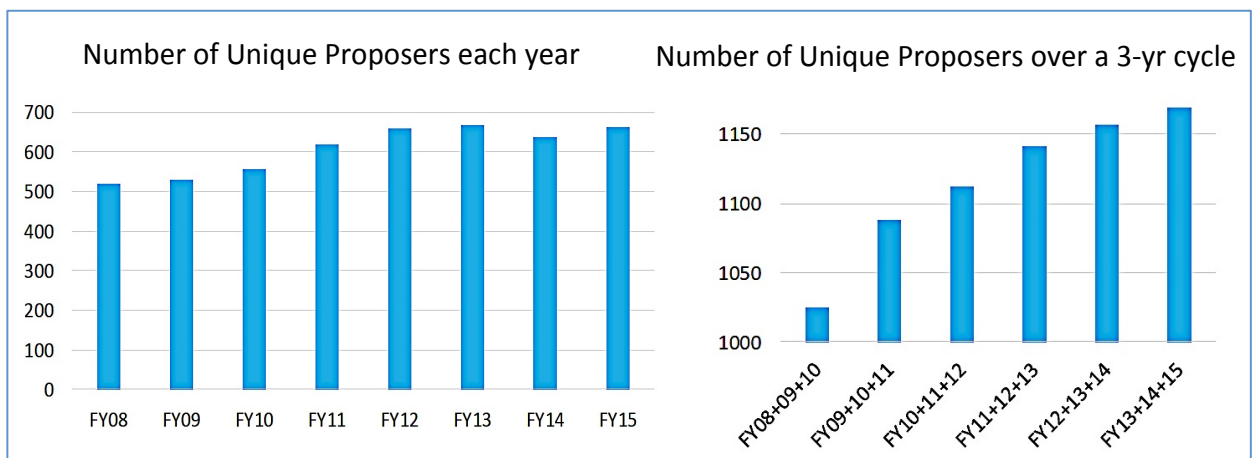

*Figure 2. Trending plots showing the number of unique individuals submitting to NSF/AST AAG program as PI each year, as well as the sum over 3 years corresponding to a typical grant cycle. Declined proposals can be re-submitted the next year, but PIs with accepted proposals will not resubmit for the same project until after (typically) 3 years.*

### 3. What should be the minimally acceptable funding rate for meritorious science?

Having established that the steady decline in proposal success rates is not being driven principally by changes in proposer demographics, nor in the merit of the science proposed, we conclude that, in the current funding state, there is untapped potential resulting in loss of valuable scientific research. The publicly-owned data products generated from first-rate astronomical assets, facilities, and missions are being under-utilized, as indicated by the large number of highly-ranked, but unfunded, proposals for their use. While investment by both federal and non-federal sources have generated a large *capacity* for new meritorious science, federal funding alone is no longer sufficient to support researchers to carry out the science opportunities thus provided. We next consider the question: What should be adopted as the minimal acceptable funding success rate? To answer this question, we draw upon and extend recent research that uses empirical data to develop a statistical model of astronomy grant proposers and success rates[3].

*3.1 Probabilities of funding success*

Table 1 uses the statistical model of von Hippel & von Hippel (2015) to consider various hypothetical scenarios of funding and resubmission rates, assuming that the typical investigator submits a single proposal to a given opportunity (consistent with real behavior as illustrated in Section 1). Scenarios corresponding to NSF Astronomy funding rates in FY2003 and those projected for FY2018 in the absence of any facility divestment are highlighted as green and red, respectively. Highlighted in yellow is the minimum benchmark for scientific productivity, defined as the point at which the time it takes new researchers to write potentially unfunded proposals "costs" more in terms of scientific output than the papers facilitated by a successful proposal (see Section 3.2). Note that the historically much higher success rate of 30-35% in FY2003 represented a healthily competitive environment, in which the average investigator faced a manageable level of risk (~30%) of no funding after three attempts (Table 1). In the current situation, where the success rate of proposals rated "very good" has dropped to ~25% (Sec. 2.2), Table 1 shows that investigators have only a ~58% probability of being funded even after three attempts.

Importantly, von Hippel & von Hippel (2015) show that, due to reviewers rating currently-funded investigators more highly (a well-known rater bias known as the Matthew Effect[10]), the average funding rate is always an over-estimate of the true rate for new researchers. For example, during 2009 to 2012, with average funding rates ranging from 18% (psychology proposals to NSF) to 28% (astronomy proposals to NASA), the von Hippel study found that researchers with current funding had a ~50% probability of being funded in the next cycle compared to only ~7% for new or recently unfunded researchers. This effect is also apparent for researchers who have not been successfully funded in recent years. For a 20% average funding rate, presently unfunded and new investigators, therefore, compete with one another for an effective funding rate of only 12%. Using these conditional probabilities for three consecutive attempts, one-eighth of the presently funded researchers $(1 - 0.5)^3$ plus two-thirds of the presently unfunded or new researchers $(1 - 0.12)^3$, or a total of ~80% of proposers, will be unable to secure grants for their research in a three-year funding cycle.

In this scenario, for a funding rate around 20%, even investigators who currently have success with funding proposals may soon find themselves in a "meta-stable" situation, in which they experience a reasonably good funding rate for some number of years before the probabilities catch up with them and they "phase transition" into the unsuccessful group, and therefore drop into the long-unfunded category.

| PROPOSAL SUCCESS RATE | P (no funding) 1 try | P (no funding) 2 tries | P (no funding) 3 tries | P (no funding) 4 tries | P (no funding) 5 tries |
|---|---|---|---|---|---|
| *10%* | 90% | 81% | 73% | 66% | 59% |
| *15%* | 85% | 72% | 61% | 52% | 44% |
| *20%* | 80% | 64% | 51% | 41% | 33% |
| *25%* | 75% | 56% | 42% | 32% | 24% |
| *30%* | 70% | 49% | 34% | 24% | 17% |
| *35%* | 65% | 42% | 27% | 18% | 12% |

*Table 1. Probabilities of unfunded proposals for different hypothetical funding rates and number of proposal attempts. The green shaded cell represents the state of the field circa 2003 (see Fig. 1). The red shaded cell represents the impending situation expected by FY2018 in the absence of portfolio rebalancing. The yellow shaded cell is the nominal "absolute minimum" benchmark identified here as the point at which new researchers spend more time proposing than publishing papers; it is not a sustainable benchmark and should be regarded as a temporary acceptable minimum.*

---

[10] Merton, R.K. 1968, Science, 159, 56.

*3.2 The cost of time and scientific productivity*

Three quarters of the proposers in the von Hippel study were at large research universities. The survey found that principal investigators spend 116 hours on average per proposal, and co-investigators spend 55 hours. This translates into more than 8 PI calendar-weeks and 3.8 Co-I calendar-weeks per proposal for the typical research-active academic, who devotes 10-15 hours per week to scholarly activity. Not only is there a time cost, but there is also a paper-publication cost. The von Hippel study found that the effort spent in writing a typical proposal costs these investigators an average of 0.41 papers per NSF astronomy proposal and 0.48 papers per NASA astronomy proposal. If the funding rate averages only 20%, the cost per successful proposal becomes 2.1 or 2.4 papers, respectively. On the other hand, the number of papers facilitated by grants in astronomy is approximately 7.9 papers per grant. If astronomy funding rates were to drop to less than ~6%, time spent on proposal writing would exceed that spent on any potential research papers.

*3.3 The cost to the Review Process*

The increased proposal load poses significant review management challenges to the funding agencies. Although agency staffing has remained relatively flat in recent years, the number of panels, each with a higher number of proposals, has increased. The organization and execution of just one panel takes 130+ hours of each NSF Program Officer's time. NSF has developed a number of tools to optimize the internal review processes, but another 30% increase in proposal volume over the next five years would not be manageable. In 2014, NASA/APD handled 832 proposals in its core R&A programs. The estimated yearly cost in NASA staff time and direct expenses for reviewer travel, meeting space, etc. to plan, execute, and document the evaluation and selection process is ~$3M.

Another issue is recruitment of reviewers. This is a larger problem for NSF/AST who do not allow an individual listed as PI or co-PI on an NSF/AST AAG proposal to serve as a reviewer. The result is that over 1,100 qualified individuals are prohibited from joining a panel. It is thus more difficult to recruit un-conflicted members of the community to join the panels. Reviewer acceptance rates have been falling; currently only 20-25% of reviewers agree to serve when asked. For NASA, conflict of interest issues can often be mitigated by putting the reviewer on a different panel. The growing number of proposals has not yet caused an increase in COI issues for NASA.

## 4. Concluding Remarks

An average funding rate of ~6% in astronomy would represent an unsustainably low rate for the health of the field, and would fall far short of US aspirations for leadership, with a vibrant program that attracts the brightest young researchers. Yet, as discussed above, this is close to the effective funding rate for new investigators or investigators who have recently transitioned into unfunded status, even for an average success rate of 20%, and the current average is already below this in NSF/AST and in NASA Heliophysics. Such a low success rate costs our field an immense amount of scientifically productive time and pushes investigators away from grant-supported research. The nation's investment in missions and state-of-the-art facilities is founded on the belief that these facilities will lead to scientists producing great science with them. We should strive for a funding success rate that prevents proposal writing costs from overtaking scientific productivity, and that encourages new researchers to pursue original, potentially transformative ideas.

While this strategic advice is supported by the data shown above, the appropriate tactical solutions required to keep astrophysics viable in the US are far from clear. With a fixed budget, the number of individual investigator awards is limited by existing or proposed commitments to space-based missions or ground-based facilities operations. In the face of constrained federal budgets, this portfolio balance becomes one of the few knobs that can be used to adjust the proportion of individual research grants that are funded. For NSF/AST, the grants-to-facilities balance improved from a value of 35:65 in the late 1990s to a value close to 50:50 around 2008. The balance is now inching back towards 40:60, with even lower ratios predicted for the future. This led to the 2012 Portfolio Review, which urged divestment of some optical and radio telescopes, but these adjustments will not be sufficient to pull the success rates above the 20% threshold. The situation in NASA Astrophysics is somewhat better at present, and there is the hope of increasing funding through upcoming Guest Observer opportunities (e.g., in connection with JWST). Over the last decade the NASA Heliophysics program has increasingly emphasized a few larger missions at the expense of smaller missions and the competed research program. The Solar and Space Physics survey gave specific recommendations to rebalance the portfolio in the next decade, with an emphasis on research and analysis embodied in the DRIVE program. NSF/GEO/AGS is currently conducting its own portfolio review of geospace facilities in order to address this challenge of operating new facilities versus the expense of existing facilities and the competed research programs.

Another adjustable knob could be the size of the average grant budget, but data indicate that proposal budgets are not growing out of line with inflation. The median proposal request for NSF/AST AAG program has increased from $93k/year to $150k/year over the last 25-year period, which corresponds to a 12% reduction in constant 2015 dollars.  If anything, reducing the size of grant budgets will exacerbate the problem, since more successful proposals are then required to support the average researcher or team. Reducing the size of individual research grants, decreasing the number of funding opportunities available, limiting the number of proposal submissions per investigator, or enforcing a pre-proposal stage may artificially raise funding rates in the short term, but will actually only serve to disguise the problem instead of solving it. Eventually it is not worth applying for a funding opportunity that cannot credibly support the science proposed.

An adjustable knob of last resort may be reduction of the pool of investigators.  Indeed ample anecdotal evidence indicates that researchers are now leaving the field at various career stages, or leaving the U.S. to pursue their scientific careers in more promising funding environments abroad. Thus, we are already well on our way toward limiting the growth of the U.S. astronomy community, but in an uncontrolled manner that wastes resources, depletes from our ranks the best and brightest thinkers, and risks losing U.S. leadership in the most crucial areas.

In summary, both the low overall funding and the ever-increasing number of meritorious proposals (and scientists they represent) are creating an unsustainable tension. We are concerned that we have reached a tipping point where the health and vitality of the community cannot be sustained. It is our hope that a data-driven response to these issues can enable the best science and preserve our leadership position in this most fundamental of sciences, even in these lean budgetary times. This grave situation warrants a fresh look at rebalancing the portfolio and taking the necessary steps to protect the grants programs.

**Acknowledgements:**  The NSF and NASA data presented in this report could not have been assembled and analyzed in a useful and timely way without the active involvement of Hashima Hasan, Linda Sparke, Daniel Evans, Arik Posner, Jonathan Rall, Jim Ulvestad, Jim Whitmore, and Paul Hertz.  We gratefully acknowledge their help and encouragement.